\documentclass[12pt,preprint]{aastex}

\begin{document}

\title{Log-Poisson Hierarchical Clustering of Cosmic Neutral Hydrogen
and Ly$\alpha$ Transmitted Flux of QSO Absorption Spectrum}

\author{Yi Lu\altaffilmark{1,2}, Yao-Quan Chu\altaffilmark{1}, and
Li-Zhi Fang\altaffilmark{2}}

\altaffiltext{1}{Center for Astrophysics, University of Science and
Technology of China, Hefei, Anhui 230026, China}
\altaffiltext{2}{Department of Physics, University of Arizona,
Tucson, AZ 85721}

\begin{abstract}

We study, in this paper, the non-Gaussian features of the mass
density field of neutral hydrogen fluid and the Ly$\alpha$
transmitted flux of QSO absorption spectrum from the point-of-view
of self-similar log-Poisson hierarchy. It has been shown recently
that, in the scale range from the onset of nonlinear evolution to
dissipation, the velocity and mass density fields of cosmic baryon
fluid are extremely well described by the She-Leveque's scaling
formula, which is due to the log-Poisson hierarchical cascade. Since the
mass density ratio between ionized hydrogen to total hydrogen is not
uniform in space, the mass density field of neutral hydrogen
component is not given by a similar mapping of total baryon fluid.
Nevertheless, we show, with hydrodynamic simulation samples of the
concordance $\Lambda$CDM universe, that the mass density field of
neutral hydrogen, is also well described by the log-Poisson
hierarchy.

We then investigate the field of Ly$\alpha$ transmitted flux of QSO
absorption spectrum. Due to redshift distortion, Ly$\alpha$ transmitted
flux fluctuations are no longer to show all features of the
log-Poisson hierarchy. However, some non-Gaussian features predicted
by the log-Poisson hierarchy are not affected by the redshift
distortion. We test these predictions with the high resolution and
high S/N data of quasars Ly$\alpha$ absorption spectra. All results
given by real data, including $\beta$-hierarchy, high order moments
and scale-scale correlation, are found to be well consistent with the
log-Poisson hierarchy. We compare the log-Poisson hierarchy with the
popular log-normal model of the Ly$\alpha$ transmitted flux. The later
is found to yield too strong non-Gaussianity at high orders, while
the log-Poisson hierarchy is in agreement with observed data.

\end{abstract}


\section{Introduction.}

Baryon matter of the universe is mainly in the form of
intergalactic medium (IGM), of which the dynamics can be described
as compressible fluid. Luminous objects are formed from baryon matter
in the gravitational well of dark matter. Therefore, the dynamical
state of baryon fluid in nonlinear regime is crucial to understand
the formation and evolution of the large scale structures of the
universe. In the linear regime, the baryon fluid follows the mass
and velocity fields of collisionless dark matter. In the nonlinear
regime, however, the baryon fluid statistically decouples from the
underlying dark matter field. The statistical behavior of baryon
fluid is no longer described by a similar mapping of the underlying
dark matter field (e.g. Pando et al. 2004).

It was first pointed out by Shandarin and Zeldovich (1989) that the
dynamical behavior of baryon matter clustering on large scales is
similar to turbulence. The expansion of the universe eliminates
the gravity of uniformly distributed dark matter. The motion of baryon
matter on scales larger than dissipation is like that of matter moving
by inertia. In this regime, the evolution of baryon matter is scale-free
and dynamically like fully developed turbulence in inertial range.
The turbulence of incompressible fluid leads the energy passes from
large to the smallest eddies, and finally dissipates into thermal motion.
While the clustering of cosmic baryon fluid is also due to the transform
of density perturbations on different scales, and finally falls and
dissipates into virialized halos of dark matter. Yet, the turbulence of
incompressible fluid is rotational (Landau \& Lifshitz 1987), while the
clustering of cosmic matter is irrotational, because vorticities
do not grow in an expanding universe (Peebles 1980).

Nevertheless, the turbulence-like behavior of cosmic baryon fluid
has been gradually noticed. First, the dynamics of growth modes of
the cosmic matter is found to be sketched by a stochastic force
driven by Burger's equation (Gurbatov, et al 1989, Berera \& Fang,
1994). The Burger's equation driven by the random force of the
gravity of dark matter can also sketch the evolution of baryon
fluid, if cooling and heating are ignored (Jones 1999; Matarrese \&
Mohayaee 2002). Later, the Burger's fluid is found to show
turbulence behavior if the Reynolds number is large enough (Polyakov
1995; Lassig 2000; Bec \& Frisch 2000; Davoudi et al. 2001). The
Reynolds number of IGM at nonlinear regime actually is large.
Therefore, we may expect that, in the scale free range, the
dynamical state of cosmic baryon fluid should be Burger's
turbulence. The turbulence of Burger's fluid is different from the
turbulence of incompressible fluid. The later consists of vortices
on various scales, while the former is a collection of shocks.

With the cosmological hydrodynamic simulation based on Navier-Stokes
equations in which heating and cooling processes are properly
accounted, it has been found that the velocity field of the IGM consists of
an ensemble of shocks, and satisfies some scaling relations predicted by
Burger's turbulence (Kim et al. 2005). This result reveals that the
turbulence features of cosmic baryon fluid are independent of the
details of dissipation (heating and cooling) mechanism if we consider only
the scale free range, i.e. from the scale of the onset of nonlinear evolution to
the scale of dissipation, say Jeans length.

A new progress is to show that the velocity field of cosmic baryon fluid can
extremely well described by She-Leveque's (SL) scaling formula (He et al.
2006). The SL formula is considered to be the basic statistical features of the
self-similar evolution of fully developed turbulence. Very recently, the
non-Gaussianities of mass density field of the hydrodynamic simulation
samples are found to be well consistent with the predictions of the
log-Poisson hierarchy, which originates from some hidden symmetry of the
Navier-Stokes equations. This hierarchical model gives a unified explanation
of non-Gaussian features of baryon fluid, including the intermittence,
hierarchical relation, scale-scale correlations etc (Liu \& Fang
2008). These results strongly indicate that, in the scale free range, dynamical
state of cosmic baryon fluid is similar to a fully developed turbulence.

In this paper, we investigate the log-Poisson hierarchy of cosmic
baryon fluid with observed data -- the Ly$\alpha$ transmitted flux
of quasar absorption spectrum, which is due to the absorptions of
quasar continuum by the diffusely distributed neutral hydrogen (Bi
et al 1995, Bi \& Devidsen 1997, Rauch 1998). These samples offer a
unique way to study the non-Gaussian feature of cosmic baryon fluid.
It has been known for a long time that the fields of Ly$\alpha$
forests and transmitted flux are highly non-Gaussian. Observation
samples of Ly$\alpha$ forests and transmitted flux show scale-scale
correlation (Pando et al 1998), intermittence (Jamkhedkar et al.
2000; Pando et al. 2002; Feng et al. 2003), non-thermal broadening
of \ion{H}{1} and \ion{He}{2} Ly$\alpha$ absorption lines (Zheng et
al. 2004; Liu et al. 2006). We will show that the log-Poisson
hierarchy provides a crux to understand the non-Gaussian behavior.

The outline of this paper is as follows. \S 2 gives an introduction of the
log-Poisson hierarchy. \S 3 shows that the neutral hydrogen component of cosmic
baryon fluid is of log-Poisson hierarchy. In \S 4, we study the log-Poisson
hierarchy of the field of Ly$\alpha$ transmitted flux with observed samples of quasar
absorption spectra. A comparison between log-Poisson hierarchy and
log-normal model is also presented in \S 4. The conclusion and discussion are
given in \S 5.

\section{Log-Poisson hierarchy}

\subsection{Structure function}

To describe the statistical properties of an isotropic and homogenous random field
$\rho({\bf x})$, it generally uses correlation functions of
$\delta \rho({\bf x})=\rho({\bf x})-\bar{\rho}$, $\bar{\rho}$ being the mean of
density. For instance, a two-point correlation function is
$\langle \delta \rho({\bf x})\delta \rho({\bf x'})\rangle$. To reveal the
turbulence-like
behavior, we use variable $\delta\rho_{r}= \rho({\bf x+r})-\rho({\bf x})$,
where $r=|\bf r|$. The variable $\delta\rho_{r}$ is very different from variable
$\delta \rho({\bf x})$. The later can be larger than $\bar{\rho}$, but cannot be less
than $-\bar{\rho}$, and therefore, for a nonlinear field, the distribution of $\delta
\rho({\bf x})$ generally is skew; however, the distribution of $\delta\rho_{r}$ is
symmetric with respect to positive and negative $\delta\rho_{r}$ if the field is
statistically uniform.

With $\delta \rho({\bf x})$ the clustering and non-Gaussianity of mass density
$\rho({\bf x})$ are measured by two and multiple point correlation functions of
$\delta \rho({\bf x})$. With $\delta \rho_{r}$, the statistical features are
described by structure function defined as
\begin{equation}
S_p(r)\equiv \langle |\delta\rho_r|^p\rangle,
\end{equation}
where $p$ is the order of statistics, and the average $\langle ...\rangle$
is taken over the ensemble of density field. A comparison of the correlation
function and structure function has been analyzed in detail by Monin \& Yaglom
(1975). The 2nd structure function $S_2=\langle |\delta\rho_r|^2\rangle$ as a
function $r$ (scale) actually is the power spectrum of the mass density field
$\rho({\bf r})$ (Fang \& Feng 2000).

In the scale-free range of the dynamical equations and initial conditions,
the structure function is scale-invariant, and therefore, it is generally
expressed as a power law of $r$
\begin{equation}
S_p(r)\propto r^{\xi(p)}.
\end{equation}
$\xi(p)$ is called intermittent exponent. Since the pioneer work of
Kolmogorov (1941), it is believed that the relation of $\xi(p)$ vs.
$p$ is related to the scale-covariance of the dynamical equations
and initial conditions. For fully developed turbulence of
Navier-Stokes fluid, $\xi(p)$ is a nonlinear function of $p$. Since
then many hierarchy models for interpreting $\xi(p)$ have been proposed
(Frisch 1995). Finally the best model is given by the SL scaling
formula (She \& Leveque 1994), which is yielded from the Log-Poisson
hierarchy process (Dubrulle 1994).

Although the cosmic baryon fluid is not incompressible, samples of
mass and velocity fields of cosmic baryon fluid produced by the
cosmological hydroidynamic simulation of the concordance
$\Lambda$CDM model show in good arrgement with the SL scaling and
log-Poisson hierarchy. This is not surprising, because the
hierarchical structure model is mainly based on the invariance and
symmetry of nonlinear dynamical systems. Therefore, systems other
than the Navier-Stokes incompressible fluid will also show the SZ
scaling and log-Poisson hierarchy if they have the similar
invariance and symmetry (She 1997).

\subsection{Log-Poisson hierarchical cascade}

The scenario of hierarchical clustering has been widely used to
describe nonlinear evolution of the mass field of cosmic matter. We
will first give the basic assumptions of log-Poisson hierarchy
cascade, and then discuss the physics behind this model.

The log-Poisson hierarchy assumes that, in the scale free range, the
variables (density fluctuation) $|\delta\rho_{r}|$ on different scales $r$
are related by a statistical relation as (Dubrulle 1994, She \&
Waymire 1995),
\begin{equation}
|\delta\rho_{r_2}| = W_{r_1r_2}|\delta\rho_{r_1}|,
\end{equation}
where
\begin{equation}
W_{r_1r_2}=\beta^m (r_1/r_2)^{\gamma}.
\end{equation}
In eqs.(3) and (4), $r_1\geq r_2$, and $m$ is a random variable with the
Poisson PDF as
\begin{equation}
P(m)=\exp(-\lambda_{r_1r_2})\lambda_{r_1r_2}^m/m!.
\end{equation}
To insure the normalization $\langle W_{r_1r_2} \rangle=1$, where
$\langle...\rangle$ is over $m$, the mean $\lambda_{r_1r_2}$ of the
Poisson distribution should be
\begin{equation}
\lambda_{r_1r_2}= \gamma[\ln(r_1/r_2)]/(1-\beta).
\end{equation}

The model eq.(3) describes how a density fluctuation $|\delta\rho_{r_1}|$
on larger scale $r_1$ statistically related to fluctuation $|\delta\rho_{r_2}|$
on smaller
scale $r_2$.  The log-Poisson model depends only on the ratio
$r_1/r_2$. Thus, it is scale invariant. The random variable $m$ can be
considered as the steps of the evolution from $|\delta\rho_{r_1}|$ to
$|\delta\rho_{r_2}|$. When $\beta$ is smaller, only the evolutionary
process with lower steps is important.

For a Gaussian field, variables $\delta\rho_{r_1}$ and
$\delta\rho_{r_2}$ have to be statistically independent. It likes
that the Fourier modes with different wavenumber $k_1\propto 1/r_1$
and $k_2\propto 1/r_2$ are statistically independent. Therefore, a
Gaussian field has to be $\beta = 1$. Thus, the parameters $\beta<1$
is a measure of the deviation from Gaussian field. The meaning of
$\gamma$ will be given later.

Among hierarchical clustering models, the log-Poisson hierarchy has
the following features. First, the relation between
$|\delta\rho_{r_1}|$ and $|\delta\rho_{r_2}|$ given by eq.(3) is a
multiplicative random process. A random multiplicative cascade
generally yields a non-Gaussian field. That is, even the field on
large scale $r_1$ is Gaussian, it will be non-Gaussian on small
scale $r_2$. This is different from additive random process (e.g.
Cole \& Kaiser 1988), which generally yields Gaussian field (Pando
et al. 1998).

Second, the cascade from scale $r_1$ to $r_2$, and then to $r_3$ is
identical to the cascade from $r_1$ to $r_3$. It is because
$W_{r_1r_3}=W_{r_1r_2}W_{r_2r_3}=\beta^N (r_1/r_3)^{\gamma}$, where
$N$ is again a Poisson random variable with
$\lambda_{r_1r_3}=\lambda_{r_1r_2}+\lambda_{r_2r_3}$. Therefore, the
log-Poisson hierarchy removes an arbitrariness in defining the steps
of cascade from $r_1$ to $r_2$ or $r_2$ to $r_3$. This arbitrariness
is a major shortcoming of many hierarchical clustering
models, one of them,  for instance, is the clustering hierarchy
models (Soneira \& Peebles 1977, Peebles 1980).

Third, although the log-Poisson hierarchical process is discrete in
terms of the discrete random number $m$, it is infinitely divisible.
That is, there is no lower limit on the difference  $r_1-r_2$. It
can be infinitesimal. This is consistent with the continuous
variable $r$ used in the hydrodynamic equation of cosmic baryon
fluid. The infinite divisibility can not be modeled with hierarchy
of discrete objects with finite size.  Log-normal model is also
infinitely divisible. However, their asymptotic behavior of $\xi(p)$
at larger $p$ is unbound. We will compare the log-normal model with
log-Poisson in \S 4.4.

\subsection{Intermitted exponent}

With log-Poisson hierarchy eq.(3), the intermittent exponent $\xi(p)$
is found to be (Liu \& Fang 2008)
\begin{equation}
\xi(p)=-\gamma[p-(1-\beta^{p})/(1-\beta)].
\end{equation}
When $\beta \rightarrow 1$, we have $\xi(p)=0$. Therefore, $\beta= 1$ is a
Gaussian field. A field with $\beta<1$ is called intermittent. Eq.(7) requires
$\xi(1)=0$. Therefore, the difference of an intermittent field from
Gaussian is mainly given by the term $\gamma(1-\beta^{p})/(1-\beta)$ of
eq.(7).

From eq.(7), the power spectrum $S_2(r)={\rm const}$ is
flat. This is not generally applicable for the cosmic density field.
In the scale free range, the power spectrum of mass field is of
power-law. Thus, we should generalize the
log-Poisson hierarchy eq.(3) by replacing $\delta\rho_{r_1}$ and
$\delta\rho_{r_2}$ with $r_1^{\alpha}\delta\rho_{r_1}$ and
$r_2^{\alpha}\delta\rho_{r_2}$. In this case, the intermitted exponent
$\xi(p)$ is
\begin{equation}
\xi(p)=-\alpha p-\gamma[p-(1-\beta^{p})/(1-\beta)].
\end{equation}
Eqs.(2) and (7) yield $S_2(r)\propto r^{-2\alpha}$, and therefore, the
parameter $2\alpha$ is the index of power spectrum.  When  $\alpha=0$,
eq.(8) is the same as eq.(7).

\subsection{$\beta$-hierarchy}

Since $|\delta\rho_r|^p$ is the $p$-th moment of the $\delta\rho_r$,
for high $p$, one can attribute the structure function $S_p(r)$ to
the events located at the tail of the probability
distribution function (PDF) of $\delta\rho_r$. To pick up the
structures, which dominant the $p$-order statistics of
$\delta\rho_r$, we define
\begin{equation}
F_p(r) \equiv S_{p+1}(r)/S_p(r).
\end{equation}
From equations (2) and (8), equation (9) gives
\begin{equation}
F_p(r) =A r^{-\alpha -\gamma(1-\beta^p)}.
\end{equation}
where the constant $A$ is independent of $r$ and $p$. For an
intermittent field $\beta<1$, we have $F_{\infty}(r) =A
r^{-\alpha-\gamma}$. Thus, from eq.(10), one can find
\begin{equation}
\frac{F_{p}(r)}{F_{\infty}(r)}= \left
[\frac{F_{p+1}(r)}{F_{\infty}(r)}\right]^{1/\beta},
\end{equation}

Equation (11) is invariant with respect to a translation in $p$.
Since $F_p(r)$ measures the structures dominating the $p$ order
statistics, the larger the $p$, the larger the contribution of
strong-clustered structures to $F_p(r)$. Therefore, equation (11)
describes the hierarchical relation between the stronger (or high
$p$) and weaker (or low $p$) clustering. In the scale-free range
where $F_{p+1}(r)/F_{\infty}(r)<1$, we have $F_{p}(r)/F_{\infty}(r)<
F_{p+1}(r)/F_{\infty}(r)$ if $\beta<1$. That is, for an intermittent
field, weak clustering structures are strongly suppressed with
respect to the strong clustering; the smaller the $\beta$, the
stronger the suppression of weak clustering structures.

From eq.(11), we have
\begin{equation}
\ln F_{p+1}(r)/F_3(r)=\beta\ln F_{p}(r)/ F_2(r).
\end{equation}
This, for {\it all} $r$ and $p$, $\ln [F_{p+1}(r)/F_3(r)]$ vs.
$\ln [F_{p}(r)/F_2(r)]$ should be on a straight line with slope
$\beta$. It is called $\beta$-hierarchy. Eq.(12) does not contain
parameter $\gamma$ and term $F_{\infty}(r)$, and therefore, it is a
testable prediction of the log-Poisson hierarchy.

Therefore, log-Poisson hierarchy links the sizes of fluctuation
structures [eq.(3)] as well as their amplitude or intensity [eq.(11)].
This is different from hierarchy models, which give only
the relationship between the sizes of objects. The hierarchical relation
eq.(11) actually is the origin of the log-Poisson hierarchy. That is,
the hierarchical relation like eq.(11) is the first recognized to be held
for dynamical system described by the Naiver-Stokes equations, or equations
close to the Naiver-Stokes fluid (Dubrulle 1994, Leveque \& She 1997).
Therefore, the SL formula has also been successfully applied to describe the
mass fields of compressible fluid (Boldyrev et al 2002, Padoan et al 2003).

\subsection{$\gamma$-related non-Gaussianities}

The ratio between higher order to second order moments $\langle
\delta\rho_r^{2p}\rangle/\langle\delta\rho_r^{2}\rangle^p$ is a
popular tool to measure non-Gaussianity. When $p=2$, the ratio is
kurtosis. For a Gaussian field, the ratio is independent of $r$, and
equal to
\begin{equation}
\frac{\langle \delta\rho_r^{2p}\rangle}{\langle
\delta\rho_r^{2}\rangle^p}= (2p-1)!!.
\end{equation}
For log-Poisson hierarchy, one can show(Liu \& Fang 2008)
\begin{equation}
\ln \frac{\langle \delta\rho_r^{2p}\rangle}{\langle
\delta\rho_r^{2}\rangle^p}= K_p\ln r +{\rm const}
\end{equation}
The moment ratio
 $\ln(\langle\delta\rho_r^{2p}\rangle/\langle\delta \rho_r^2\rangle^p)$
is linearly dependent on $\ln r$ (scale free) with the slope $K_p$
given by
\begin{equation}
K_p=-\gamma\frac{p(1-\beta^2)-(1-\beta^{2p})}{1-\beta}.
\end{equation}
As expected, for Gaussian field $(\beta\rightarrow 1)$, $K_p=0$,
i.e. the ratio of moments is independent of $r$ [Eq.(14)]. That is,
$K_p$ only depends on the last term on the right hand side of
eq.(8), regardless the parameter $\alpha$.

Other useful non-Gaussian detector is the scale-scale correlation
defined as
\begin{equation}
C^{p,p}_{r_1,r_2}\equiv \frac{\langle \delta \rho_{r_1}^{p}\delta
\rho_{r_2}^{p}\rangle} {\langle \delta \rho_{r_1}^{p}\rangle
\langle\delta \rho_{r_2}^{p}\rangle},
\end{equation}
which describes the correlation between the density fluctuations on
scales $r_1$ and $r_2$. Obviously, for a Gaussian field,
$C^{p,p}_{r_1,r_2}=1$. The clustering of cosmic large scale
structure in the nonlinear regime essentially is due to the
interaction between the modes of fluctuations on different scales
(e.g., Peebles 1980). Therefore, it is important to detect the
scale-scale correlation of cosmic baryon matter.

If the ratio $r_2/r_1$ is fixed, the log-Poisson hierarchy predicts
the scale-scale correlation to be (Liu \& Fang 2008)
\begin{equation}
C^{p,p}_{r_1,r_2}=B(r_2/r_1)r_1^{\xi(2p)-2\xi(p)},
\end{equation}
where the factor $B(r_2/r_1)$ is constant when the ratio $r_2/r_1$
is fixed. This is because the log-Poisson model is invariant with
respect to scale dilation. From eqs.(7) or (8), we have
\begin{equation}
\xi(2p)-2\xi(p)= -\gamma(1-\beta^p)^2/(1-\beta)
\end{equation}
Thus, if $r_2/r_1$ remains constant, the relationship of $\ln
C^{p,p}_{r_1,r_2}$ vs. $\ln r_1$ should be a straight line with
slope  $-\gamma(1-\beta^p)^2/(1-\beta)$. This slop is also only
dependent on the last term on the right hand side of eq.(8), regardless the
parameter $\alpha$.

\section{Log-Poisson hierarchy of neutral hydrogen density field}

Ly$\alpha$ absorption spectrum depends on the distribution of
neutral hydrogen. We first study the mass density field of diffused
neutral hydrogen. Since the UV background radiation is uniform and
does not introduce special spatial scale, one may expect that the
mass density field of neutral hydrogen is also a field of the
log-Poisson hierarchy. However, the ratio of the neutral hydrogen
density $\rho_{\rm HI}({\bf x})$ to total hydrogen density
$\rho_{\rm H}({\bf x})$ is not spatially constant, because
the temperature-density relation is of multiphase (He et
al 2004, 2005). Therefore, the mass field of the neutral hydrogen would
not have the same log-Poisson hierarchical features as of $\rho({\bf
x})$.

\subsection{Simulation samples}

The simulation samples of the fields of baryon fluid are generated
by the same way as Liu \& Fang (2008), which is based on the hybrid
hydrodynamic N-body code of Weighted Essentially Non-oscillatory
(WENO) scheme (Feng et al. 2004). It is in Eulerian scheme and
suitable to analyze the fluid in high as well as in low mass density
areas. The WENO samples have been successfully applied to reveal the
self-similar hierarchy behavior of cosmic baryon fluid (Kim et al.
2005; He et al. 2006; Liu \& Fang 2008), to explain the HI and HeII
Ly$\alpha$ absorption in quasar spectra (Liu et al. 2006), and to
study the relation between X-ray luminosity and temperature of
groups of galaxies (Zhang et al. 2006). It is also used for studying
Ly$\alpha$ leaks at high redshift (Liu et al. 2007).

The simulation is performed in a comoving cubic box of 100 $h^{-1}$
Mpc with 512$^3$ grids and an equal number of dark matter
particles. The grid size is $100/512\sim 0.2$ $h^{-1}$ Mpc, which is
less than Jeans length, and therefore, it is enough to catch the
statistical behavior in the scale-free range, i.e. larger than the
Jeans length and less than non-linear scale. We use the concordance
$\Lambda$CDM cosmology model with parameters $\Omega_{\rm m}$=0.27,
$\Omega_{\rm b}$=0.044, $\Omega_\Lambda$=0.73, $h$=0.71,
$\sigma_8$=0.84, and spectral index $n=1$. The transfer function is
calculated using CMBFAST (Seljak \& Zaldarriaga 1996). We take a
primordial composition of H and He ($X$=0.76, $Y$=0.24).

The ionization fraction is calculated with ionization-recombination
equilibrium under a uniform UV radiative background, of which the intensity is
adjusted to fit the mean of observed Ly$\alpha$ tarnsmitted flux. We
produce the distributions of the mass density of hydrogen $\rho({\bf x})$,
the fraction of neutral hydrogen $f_{\rm HI}({\bf x})$, temperature
$T({\bf x)}$, and velocity ${\bf v}({\bf x})$ at redshift $z\simeq
2.5$, which is the mean redshift of observed samples used in \S 4.

We randomly sample 10,000 one-dimensional sub-samples to simulate the
Ly$\alpha$ transmitted flux (\S 4). To estimate the errors,
we divided the 10,000 samples into 10 subsamples, each of which has 1,000
line samples.

\begin{figure}[htb]
\center
\includegraphics[scale=0.45,angle=-90]{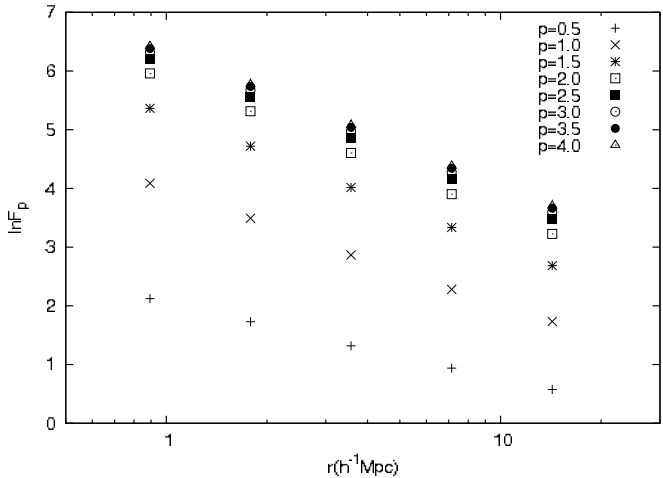}
\caption{Function $F_p(r)$ vs. $r$ of simulation samples of the mass
density field of neutral hydrogen in the scale range $0.9 < r< 15$
h$^{-1}$ Mpc, and $p=0.5\times n$, with $n=1,2...8$ from bottom to
top.}
\end{figure}

\subsection{Log-Poisson hierarchical statistics}

The variable $\delta\rho_{r}= \rho({\bf x+r})-\rho({\bf x})$ of
density field $\rho({\bf x+r})$ can be calculated by a discrete wavelet
transform (DWT) as
\begin{equation}
\delta\rho_{r,l}=\int \rho_{\rm HI}({x})\psi_{j,l}({x})d{x},
\end{equation}
where $\psi_{j,l}({x})$ is the base of discrete wavelet transform
(e.g. Fang \& Thews 1998). For a one-dimensional sample in physical
space from $x=0$ to $L$, the scale index $j$ is related to the scale
$r$ by $r=L/2^{j}$ and the position index $l$ is for the cell located
at $x=lL/2^{j}$ to $(l+1)L/2^{j}$. Because the DWT bases
$\psi_{j,l}({x})$ are orthogonal, the variables $\delta\rho_{r,l}$
do not cause false correlation. They are effective to describe
turbulence of fluid (Farge 1992). For a given scale $r$ or $j$, the
statistical average $\langle ...\rangle$ of eq.(5) is over all cells
$l$. We will use the Harr wavelet to do the calculation below. We
also repeat the calculations with wavelet Daubechies 4.

\begin{figure}[htb]
\center
\includegraphics[scale=0.45,angle=-90]{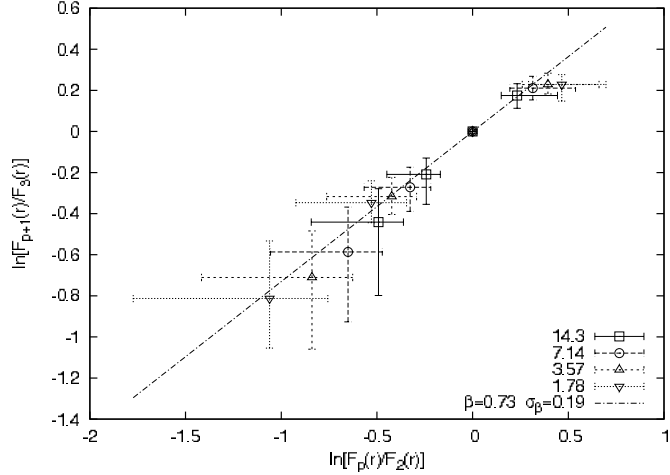}
\caption{$\ln [F_{p+1}(r)/F_3(r)]$ vs. $\ln [F_p(r)/F_2(r)]$ for
simulation samples of the mass density field of neutral hydrogen.
The data points are on scales $r=$1.78, 3.57, 7.14 and 14.3 h$^{-1}$
Mpc and the statistical order to be $p=$ 1, 1.5, 2, and 2.5.}
\end{figure}

\begin{figure}[htb]
\center
\includegraphics[scale=0.45,angle=-90]{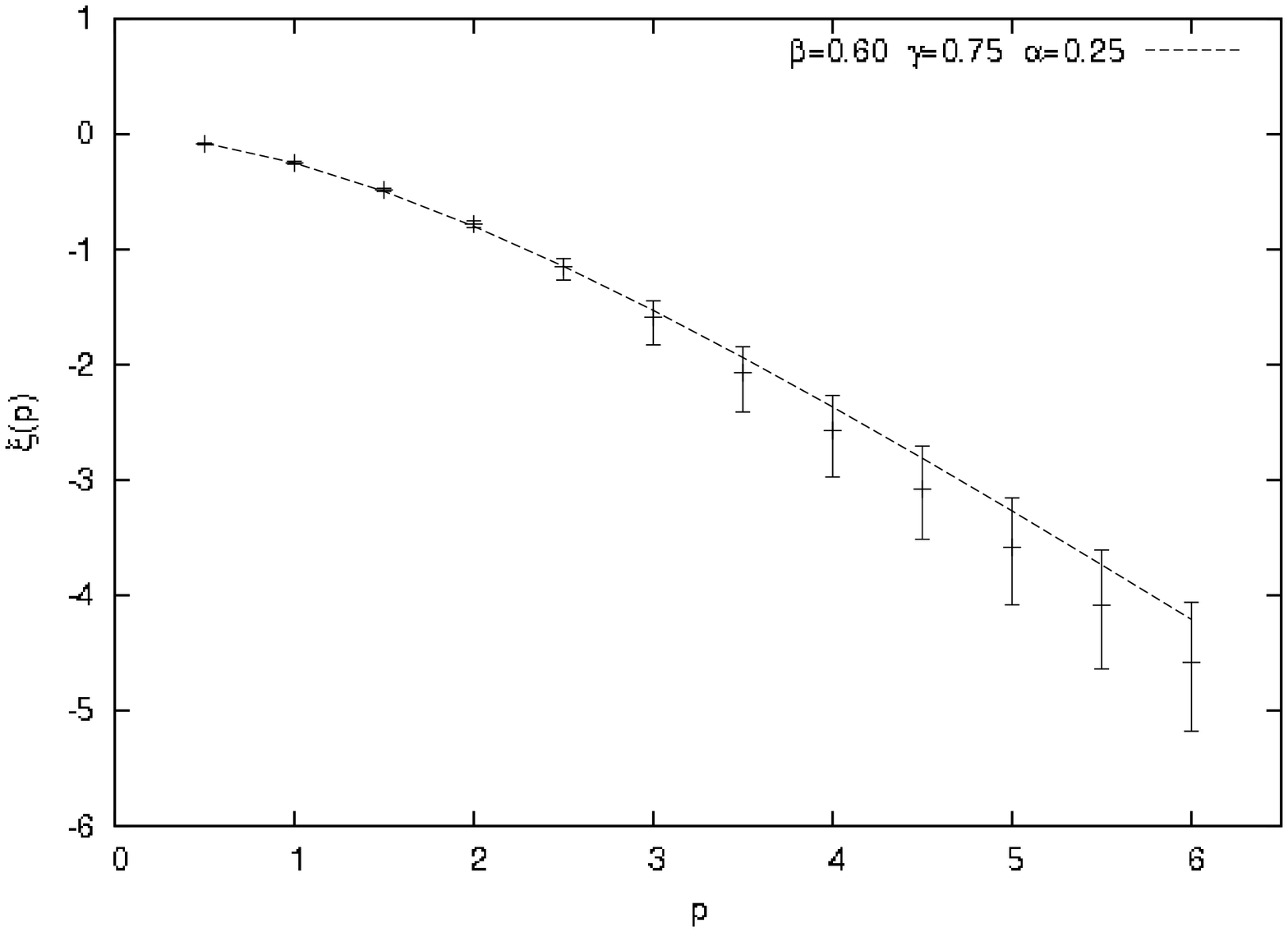}
\caption{Intermittent exponent $\xi(p)$ of simulation samples of the
mass density field of neutral hydrogen. The solid line is given by
equation (8) with parameters $\alpha=0.25$, $\beta=0.60$ and
$\gamma=0.75$. The error bars are the variance of $\xi(p)$ over 10
samples, each of which contains of 1000 one-dimensional samples.}
\end{figure}

Figures 1 - 3 demonstrate that the density field $\rho_{\rm HI}({\bf
x})$ of neutral hydrogen shows the features of log-Poisson
hierarchy. Figure 1 shows the functions $F_p(r)$ of the field
$\rho_{\rm HI}$ in the physical length scale range $0.9 <r< 15$
h$^{-1}$ Mpc and orders of $p=0.5\times n$ with $n=1, 2...8$. For
all $p$, $\ln F_p(r)$ can be well fitted by a straight line of $\ln
r$. It is consistent with eq.(10). When $p>3$, $F_p(r)$ given in
Figure 1 is almost independent of $p$. This indicates that
$F_p(r)\rightarrow F_{\infty}$ for higher $p$. Thus, from eq.(10),
$\beta$ should be less than 1, and therefore, the field is
intermittent. In this case, the slope of the straight lines with
higher $p$ has to be equal to $-\alpha-\gamma$. Figure 1 shows
$\alpha+\gamma = 1.0$.

Figure 2 is the $\beta$-hierarchy eq.(12) of the density field
$\rho_{\rm HI}$, in which the error bars are given by the ranges of
$\ln [F_{p+1}(r)/F_3(r)]$ and $\ln [F_p(r)/F_2(r)]$ of the 10 subsamples,
each of which consists of 1,000 lines. The $\beta$-hierarchy is held
for all $r$ in $1 <r< 15$ h$^{-1}$ Mpc and $p=$0.5 to 4. It yields
$\beta =0.73 \pm 0.19$.

Figure 3 plots the intermittent exponent $\xi(p)$ [eq.(8)]. It shows
that $\xi(p)$ can be well fitted with eq.(8) with parameters
$\beta=0.60$, $\alpha+\gamma=1.0$ and $\alpha=0.25$. These
parameters are in agreement with that given by Figures 1 and 2. The
error bars of Figure 3 are also the range of the 10 subsamples.
Therefore, the HI mass density distribution in the scale range of
$\sim 1$ to 15 h$^{-1}$Mpc can also be described by the log-Poisson
hierarchy. However, the parameters $\beta$ and $\gamma$ are
different from that of density field $\rho({\bf x})$, which has
$\beta \simeq 0.47$, and $\gamma\simeq 1$ (Liu \& Fang 2008). That
is, the density field $\rho_{\rm HI}({\bf x})$ is weaker
intermittent and less singular than that of $\rho({\bf x})$.

\section{Log-Poisson hierarchy of Ly$\alpha$ transmitted flux}

\subsection{Samples}

\subsubsection{Observed data}

For observed samples of Ly$\alpha$ transmitted flux, we use the high
resolution, high signal to noise ratio QSO Ly$\alpha$ absorption
spectra of Jamkhedkar (2002), Jamkhedkar et al. (2003). The power spectrum
and intermittency of this data set have been extensively and deeply analyzed
(Pando et al. 2002; Feng et al. 2002; Jamkhedkar et al.2003). It is useful
for testing the log-Poisson hierarchy.

This observational data set consists of 28 Keck High Resolution
(HIRES) QSO spectra (Kirkman \& Tytler 1997). The QSO emission
redshifts cover a redshift range from 2.19 to 4.11. The resolution
is about 8 km s$^{-1}$. For each of the 28 QSOs, the data are given
in form of pixels with wavelength, flux and noise. The noise
accounts for the Poisson fluctuations in the photon count, the noise
due to the background and the instrumentation. The continuum of each
spectrum is given by IRAF CONTINUUM fitting.

The data are divided into 12 redshift ranges from $z=1.6 + n\times
0.20$ to 1.6 + $(n+1)\times$ 0.20 where $n=0, \ldots, 11$. In our
analysis below, we use only the data in the range $z = 2.4$ to 2.6,
which  is the same as our simulation data. The scale range is taken
to be 1.54 to 12.3 h$^{-1}$ Mpc, which is also about the same as
simulation samples. The mean flux in this redshift range is
$\langle F\rangle\simeq 0.75$ (Jamkhedkar, 2002).

We use the same methods of Jamkhedkar et al. (2003) to deal with
noises, metal lines, proximity effect, and bad data chunks. On
average the S/N ratio of the Keck spectra is high. Most of the
regions with low S/N are saturated absorption regions. Although the
percentage of pixels within these regions is not large, they may
introduce large uncertainties in the analysis. We should reduce the
uncertainty given by low S/N pixels. The method is as follows.
First, we calculate the SFCs (scaling function coefficients) of both
transmission flux field $F(x)$ and noise field $n(x)$ with
$\epsilon_{jl}^F=\int F(x)\phi_{j,l}(x)dx$ and $\epsilon_{jl}^N=\int
n(x)\phi_{j,l}(x)dx$, where $\phi_{j,l}(x)$ is the scaling function
of wavelet on scale $j$ and at position $l$. We then identify
unwanted mode $(j,l)$ by using the threshold condition
$|\epsilon_{jl}^F/\epsilon_{jl}^N| < f$. This condition flags all
modes with S/N less than f.  The parameter $f$ is taken to be 3. We
also flag modes dominated by metal lines. In order to easily flag
data gaps, we set the flux at the gaps to be zero and the error to
one, and do the same thing for pixels with negative flux. Finally,
we skip all the flagged modes when doing statistics. With this
method, no rejoining and smoothing of the data are needed.

\subsubsection{Simulation samples}

To simulate Ly$\alpha$ transmitted
flux $F$, we use the 10,000 one-dimensional samples given in \S 3.1.
For each 1-D sample, the transmitted flux of Ly$\alpha$, $F(z)$ in
redshift space is calculated with $F(z)=\exp[-\tau(z)]$, where
$\tau(z)$ is the optical depth defined as
\begin{equation}
\tau(z)=\frac{\sigma_0 c}{H}\int_{-\infty}^{\infty} n_{\rm HI}(x)
V[z-x-v(x), b(x)]dx,
\end{equation}
in which $\sigma_0$ is the effective cross-section of the resonant
absorption; $H$ is Hubble constant at the redshift of the sample,
$n_{\rm HI}(x)$ is the number density of neutral hydrogen atoms. The
Voigt function is $V[z-x-v(x),
b(x)]=1/(\pi^{1/2}b)\exp\{-[z-x-v(x)]^2/b^2(x)\}$, and $b(x)$ being
the thermal broadening.  To has a proper comparison between
simulation and observation, we add noises $n_i=G \times A_i$ to each
pixel $i$, where $G$ is randomly sampled from standard normal
distribution and $A_i$ is the noise level of pixel $i$. We then
take the same data reduction of noised modes as observed data.

\subsection{Redshift distortion and $\beta$-hierarchy}

The velocity field $v(x)$ and thermal broadening $b(x)$ of eq.(20)
will lead to the deviation of the statistical properties of $\ln
F(z)=-\tau(z)$ from $n_{\rm HI}(x)$ or $\rho_{\rm HI}(x)$. The field
$\ln F(z)$ is no longer to show all features of log-Poisson
hierarchy. We should study which properties of the log-Poisson
hierarchy can still be seen with Ly$\alpha$ transmitted flux.

Because the scale free range is larger than the Jeans length, the
effect of thermal broadening would be small in this range. In this
case, eq.(20) can be approximately as
\begin{equation}
-\ln F(z)=\tau(z)=\frac{\sigma_0 c}{H}\int n_{\rm HI}(x)\delta[z-x-v(x)]dx.
\end{equation}
This relation actually is a mapping from a physical space field
$n_{\rm HI}(x)$ to redshift space field $\tau(z)$, which is the same
as that used in the redshift distortion of galaxy distribution. It
has been shown that, with the DWT variables, the redshift distortion
of eq.(21) can be estimated by (Yang et al 2002)
\begin{equation}
\delta \tau_{r,l}=\mathfrak{R}_r \delta \rho_{r,l},
\end{equation}
where the DWT variables $\delta \tau_{r,l}$ are given by $\delta \tau_{r,l}=\int
\tau({x})\psi_{j,l}({x})d{x}= \int[-\ln F(x)]\psi_{j,l}({x})d{x}$.
The redshift distortion factor $\mathfrak{R}_r$ depends on the DWT
power spectrum of the velocity field $v(x)$ on scale $r=L/2^j$.

\begin{figure}[htb]
\center
\includegraphics[scale=0.45,angle=-90]{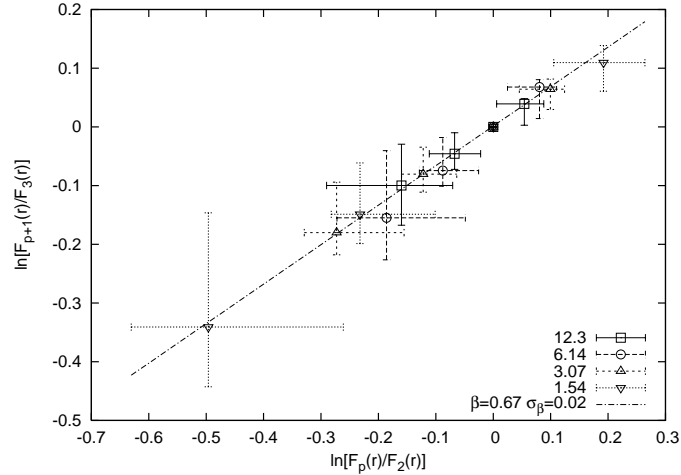}
\caption{The $\beta$-hierarchy of observed sample of the Ly$\alpha$
transmitted flux at redshift $z=2.4 - 2.6$ and physical scale range
from $\sim$ 1 to 15 h$^{-1}$ Mpc. The error bars are given by the
maximum and minimum of bootstrap resampling.}
\end{figure}

Because the average of $\langle ...\rangle$ in the structure
function eq.(1), or $S_p(r=L/2^j)=\langle |\delta\rho_{j,l}|^p\rangle$,
is only over on modes $l$, the redshift distortion factor
$\mathfrak{R}_r$ does not involve in this average. Thus, the function
$F_p(r)$ of $\delta \tau_{r,l}$ will be different from the function
$F_p(r)$ of $\delta \rho_{r,l}$ by a factor $\mathfrak{R}_r$. Thus,
both $ F_{p+1}(r)/F_3(r)$ and $F_{p}(r)/ F_2(r)$ of $\delta
\tau_{r,l}$ do not contain the redshift distortion factor
$\mathfrak{R}_r$. Thus, $F_{p+1}(r)/F_3(r)$ and $F_{p}(r)/ F_2(r)$
of $\delta \tau_{r,l}$ should also satisfy the $\beta$-hierarchy
eq.(12) as that of field $\rho_{\rm HI}$. That is, the
$\beta$-hierarchy is not affected by the redshift distortion.

Figure 4 presents the $\beta$-hierarchy of observed transmitted flux
$F(z)$. It is very well fitted by a straight line for data on
the scale range from $\sim$ 1 to 15 h$^{-1}$ Mpc. It yields
$\beta=0.67\pm 0.02$. Figure 5 shows the $\beta$-hierarchy of
observed transmitted flux, which is the same as Fig.4, and the
noised simulation samples $F$. We see that the both real and simulation
samples are well coincident. The simulation samples yield
$\beta=0.66\pm 0.02$. Therefore, both real and simulation samples are well
$\beta$-hierarchical, and the numbers of $\beta$ are well
consistent.

\begin{figure}[htb]
\center
\includegraphics[scale=0.45,angle=-90]{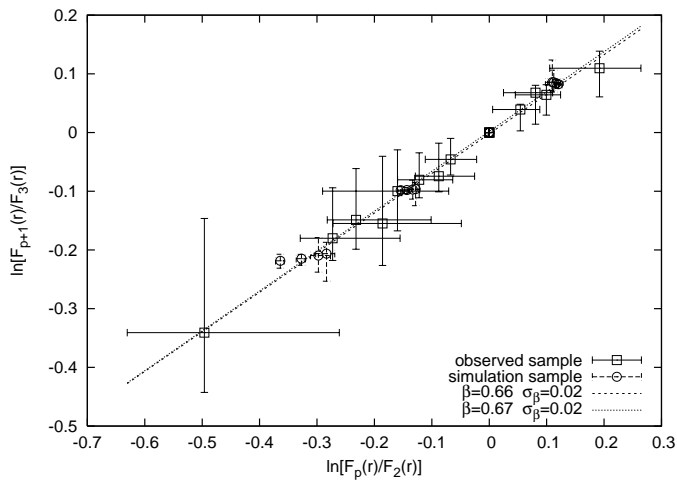}
\caption{The $\beta$-hierarchy for a. observed data (square) as Fig.
4; b. simulation sample (circle) of the Ly$\alpha$ transmitted flux
at redshift $z=2.5$ and Gaussian noises are added with the same
level as real data.}
\end{figure}

The data points of Figure 4 are scattered along the straight line
$F_{p+1}(r)/F_3(r)$-$F_{p}(r)/ F_2(r)$, while the points of Figure 5
are clustered. This is due to all simulation samples have the same
length, while for real samples consisting quasars with different
redshift, we take only the sections of transmitted flux, which are
in the redshift range $z\simeq 2.4$ - 2.6.

\subsection{High order moments}

The statistics of eq.(14) is based on the ratio between high order
moment $\langle\delta\rho_r^{2p}\rangle$ and
$\langle\delta\rho_r^{2}\rangle^p$, both of which have the same
order of $\delta\rho_r$. When $p=2$, the ratio actually is kurtosis.
The moment ratio of $\langle\delta\tau_r^{2p}\rangle$ to
$\langle\delta\tau_r^{2}\rangle^p$, obviously, is independent of the
redshift distortion factor $\mathfrak{R}_r$. Therefore, the ratio of
$\langle\delta\tau_r^{2p}\rangle$ to
$\langle\delta\tau_r^{2}\rangle^p$ has to satisfy the same property
as $\langle\delta\rho_r^{2p}\rangle$ to
$\langle\delta\rho_r^{2}\rangle^p$. Thus, for a given $p$, the
relation of
$\ln\langle(\delta\tau_r)^{2p}\rangle/\langle(\delta\tau_r)^{2}\rangle^p$
and $\ln r$ has to be a straight line with slope
$-\gamma[p(1-\beta^2)-(1-\beta^{2p})]/(1-\beta)$ [eq.(15)]. Since
the parameter $\beta$ is already determined by the
$\beta$-hierarchy, the test here is whether we can find one parameter
$\gamma$ to fit the slops of
$\langle\delta\tau_r^{2p}\rangle/\langle\delta\tau_r^{2}\rangle^p$ vs.
$\ln r$ for different $p$.

\begin{figure}[htb]
\center
\includegraphics[scale=0.45,angle=-90]{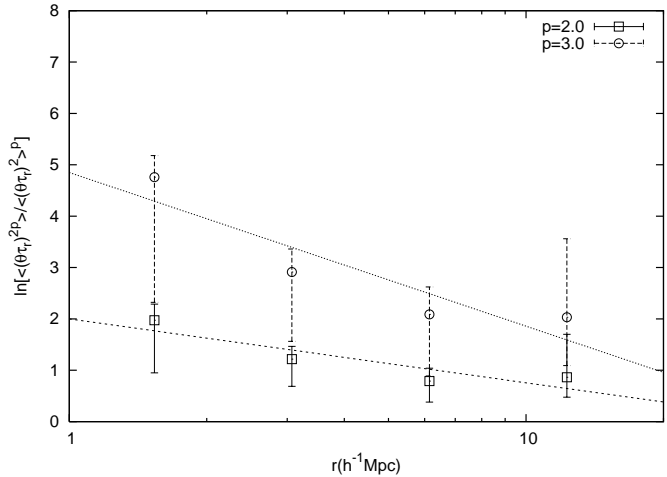}
\caption{$\ln
[\langle(\delta\tau_r)^{2p}\rangle/\langle(\delta\tau_r)^{2}\rangle^p]$
vs. $r$ of real data in the scale range $\sim 1 $ - 15 h$^{-1}$ Mpc,
and $p$ is taken to be 2 (bottom) and 3 (top). The solid lines are
given by the least square fitting, which yield slopes $0.54\pm 0.19
$ ($p=2$), and $1.3\pm 0.4$ ($p=3$). The error bars are given by the
maximum and minimum of bootstrap resampling.}
\end{figure}

The result is presented in Fig. 6. It shows the relation of $\ln (\langle
\delta\tau_r^{2p}\rangle/\langle (\delta\tau_r)^{2}\rangle^p)$ vs.
$\ln r$ for real data with $p=2$ and 3, i.e. the statistics are of
the order of 4 and 6. The slopes of the fitted straight lines are
 $0.54\pm 0.19$ for $p=2$, and $1.3\pm 0.4$ for $p=3$.
Thus, using $\beta=0.67$, we have $\gamma= 0.59\pm 0.20$ for $p=2$ and
$\gamma=0.58\pm 0.22$ for $p=3$. In spite that the statistical errors
are still large, one can already see that different $p$-lines yield
about the same parameter $\gamma$. In other words, the moment
statistics passed the test of log-Poisson hierarchy.

The number of $\gamma$ given by the real data of the transmitted
flux is less than the number $\gamma=0.75$ of $\rho_{\rm HI}$ field
(\S 3.2). Therefore, the field of transmitted flux of real samples
is less singular than the field $\rho_{\rm HI}$. This is reasonable
if considering the real data are noised.

\subsection{Log-Poisson hierarchy and log-normal model}

In \S 4.2 and 4.3., we have found that the log-Poisson parameters of
the transmitted flux should be $\beta=0.67\pm 0.02 $ and
$\gamma=0.58\pm 0.22$. This result can be used to compare the
log-Poisson hierarchy and log-normal model. We now consider the
moment ratio $\ln [\langle
\delta\tau_r^{2p}\rangle/\langle (\delta\tau_r)^{2}\rangle^p]$ as a
function of $p$ on a given scale $r$.

\begin{figure}[htb]
\center
\includegraphics[scale=0.45,angle=-90]{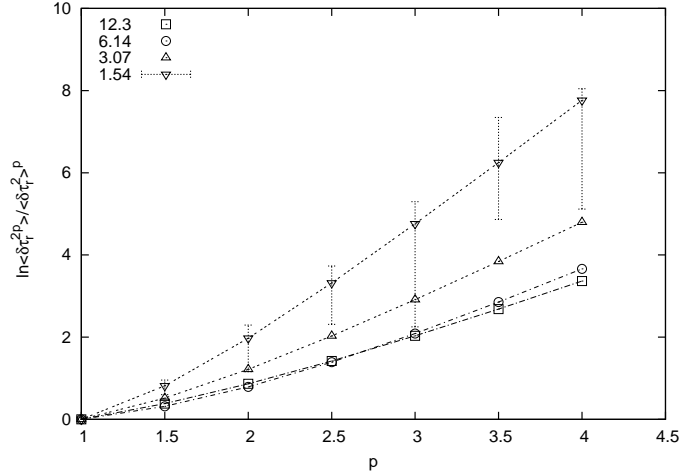}
\caption{$\ln
\langle\delta\tau_r^{2p}\rangle/\langle\delta\tau_r^{2}\rangle^p$ as
a function of $p$ for real data of Ly$\alpha$ transmitted flux at
the range $z\sim 2.4$ - 2.6 on physical scales $r=1.54$ (nabla),
3.07 (triangle), 6.14 (circle)and 12.3 (square) h$^{-1}$ Mpc.  The
curves are from the fitting of log-Poisson model. For clarify, we show
only the error bars for data points of $r=1.54$  h$^{-1}$ Mpc. The errors
of other $r$ are about the same level as $r=1.54$  h$^{-1}$ Mpc.}
\end{figure}

Figure 7 presents the $\langle \delta\tau_r^{2p}\rangle/\langle
\delta\tau_r^{2}\rangle^p$ as a function of $p$ for real data on
scales $r=1.54$ (nabla), 3.07 (triangle), 6.14 (circle)and 12.3
(square) h$^{-1}$ Mpc. $p$ is from 1 to 4. The $p$-dependent
curves shown in Figure 7 are given by eqs.(14) and (15), in which
the fitted parameters are $\beta=0.67\pm 0.02 $ and $\gamma=0.55\pm
0.10$ The $const$ in eq.(14) is determined by $\ln[\langle
\delta\tau_r^{2p}\rangle/\langle\delta\tau_r^{2}\rangle^p]=0$ when
$p=1$.

Although the observed data points of Figure 7 actually are the same 
as Figure 6, we see that the observed data points show large scatter in 
Figure 6, but almost no scatter in Figure 7. This is because Figure 6 gives
moment ratio as a $r$ function, while Figure 7 shows the $p$-dependence of 
the moment ratio. For a given $r$, the Gaussian noise will yield a moment
ratio given by eq.(13), which is a smooth function of $p$. Therefore, a 
Gaussian noisy sample should not cause scatter with respect to $p$. On 
the other hand, Gaussian noise generally is not a smooth function of $r$. 
It yields the scatter of Fig. 6.

Figure 7 shows that the $p$-dependence of moment ratio $\ln
[\langle\delta\tau_r^{2p}\rangle/\langle\delta\tau_r^{2}\rangle^p]$
is significantly dependent on scale $r$. Therefore, it can not be
fitted by a Gaussian field, for which moment ratio is
$r$-independent [eq.(13)]. The field of $\delta \tau$ is
non-Gaussian. The smaller the scale $r$, the stronger the
non-Gaussianity.

\begin{figure}[htb]
\center
\includegraphics[scale=0.45,angle=-90]{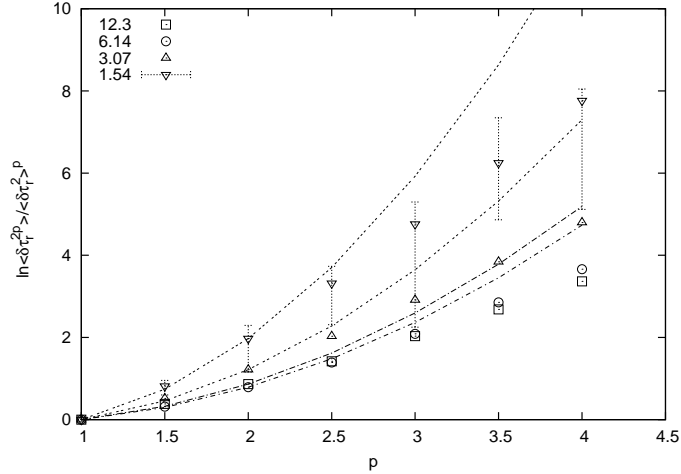}
\caption{The same as Figure 7, but the curves are from the fitting
of log-normal model.}
\end{figure}

The log-normal model of the transmitted flux of QSO Ly$\alpha$
absorption spectrum is very successful to explain various low-order
statistical features of Ly$\alpha$ forests (Bi 1993; Bi \& Davidsen
1997). The log-normal model also predicts that the transmitted flux
is non-Gaussian and intermittent. For the moment ratio, the
log-normal model yields (Pando et al 2002)
\begin{equation}
\frac{\langle \delta\tau_r^{2p}\rangle}{\langle
\delta\tau_r^{2}\rangle^p}= \exp[2(p^2-p)\sigma^2(r)]
\end{equation}
where $\sigma^2(r)$ is the power spectrum of the field. For each
$r$, one can fit eq.(24) to observed points with  $\sigma^2(r)$. The
results are plotted in Figure 8. It shows that, if we try to give a
good fitting of eq.(24) with real data at orders $p\leq 2$, the
$p$-curves of eq.(24) always give a large deviation from the real
data at $p> 2$. This deviation cannot be reduced with selecting
$\sigma^2(r)$. This is because the deviation is from the
$p^2$-dependence of $\langle \delta\tau_r^{2p}\rangle/\langle
\delta\tau_r^{2}\rangle^p$ when $p$ is large. The $p^2$-dependence
cannot be reduced with the parameter $\sigma^2(r)$.

On the other hand, at high $p$, the log-Poisson model gives $\langle
\delta\tau_r^{2p}\rangle/\langle \delta\tau_r^{2}\rangle^p \propto
p$. The increasing with $p$ is then consistent with observation. In
Figs. 7 and 8 we show the error bars for data points of $r=1.54$
h$^{-1}$ Mpc. Although the errors are large at high $p$, the result
is clearly consistent with $p$-dependence, and unfavor the
$p^2$-dependence. Therefore, the higher order statistics of the
Ly$\alpha$ transmitted flux is effective to discriminate between the
log-Poisson and the log-normal model. The log-normal model yields too
strong non-Gaussianity. This point actually has already been
mentioned in the study of turbulence (e.g. Frisch 1995).

\subsection{Scale-scale correlation}

The last statistical feature used to test the log-Poisson hierarchy is the
scale-scale correction. Similar to the statistics of high order
moment (\S 4.3), the ratio of scale-scale correction, $\langle
\delta \tau_{r_1}^{p}\delta \tau_{r_2}^{p}\rangle/\langle \delta
\tau_{r_1}^{p}\rangle \langle\delta \tau_{r_2}^{p}\rangle$ is
independent of the redshift distortion factor $\mathfrak{R}_r$. The
field $\tau$ should show the scale-scale correlation as $\rho_{\rm
HI}$ eqs.(17)-(19).
\begin{figure}[htb]
\center
\includegraphics[scale=0.45,angle=-90]{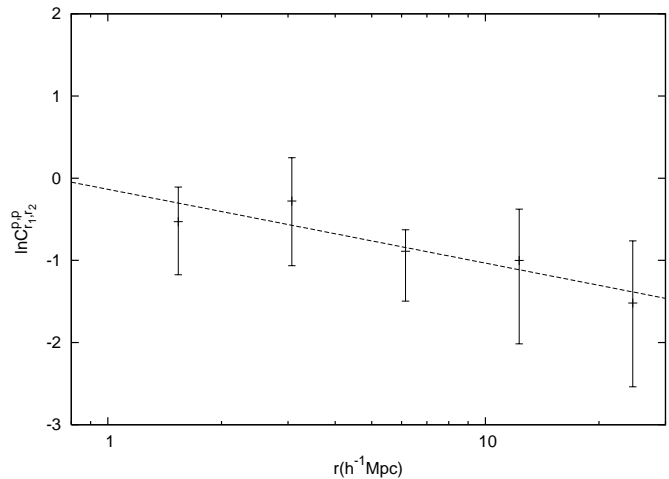}
\caption{Scale-scale correlation of observed data of Ly$\alpha$
transmitted flux at the range $z\sim 2.4$ - 2.6 for $p=2$ and
$r_2/r_1=2$. The error bars are given by the maximum and minimum of
bootstrap resampling.}
\end{figure}

Before doing this test, we would emphasize that the scale-scale
correlation is statistically independent of the statistics of high
order moments. For instance, one can construct a field, which shows
Gaussian distribution in terms of its one point distribution of
$\delta\rho_{j,l}$, but highly scale-scale correlated between
variables $\delta\rho_{j,l}$ with different $j$ (Pando et al 1998,
Feng \& Fang 2000). The clustering of cosmic large scale structure
in the nonlinear regime essentially is due to the interaction
between Fourier modes on different scales.
Therefore, cosmic clustering will definitely leads to scale-scale
correlation. Scale-scale correlation is also very effective to
distinguish various hierarchy cascade models (Pando et al. 1998).

Figure 9 presents the $p=2$ scale-scale correlation of observed
data, in which the ratio $r_2/r_1$ is fixed to be equal to 2. The
slop of the fitting straight line in Fig. 9 should be given by
eq.(19). Since eq.(19) depends only on the parameters $\beta$ and
$\gamma$, both of which have already been determined in \S 4.2 and
4.3. There is no free parameters in the fitting of Fig. 9. If we
doing a straight line, we found the parameters $\beta=0.67$ and
$\gamma=0.43\pm 0.12$. The value of $\beta$ is the same as that of
\S 4.2, while the value of $\gamma$ is smaller than that of \S 4.3,
but the deviation is not larger than 1-$\sigma$. The scale-scale
correlation is more sensitive to the quality of the data, as it is
the correlation between modes on different scales. The result of
Figure 9 is basically consistent with the scale-scale correlation
predicted by the log-Poisson hierarchy.

\section{Discussion and conclusion}

Nonlinear evolution of mass and velocity fields is a central problem
of large scale structure of the universe.  The clustering of the
cosmic baryon fluid, governed by the Navier-Stokes equation in
gravitational field of an expanding universe, has to be self
similarly hierarchical in the scale free range in which the
dynamical equations and initial perturbations are scale-invariant.

The log-Poisson hierarchical clustering sketches the nonlinear
evolution of cosmic baryon fluid in the scale free range. If the
initial density perturbations are Gaussian, and their power spectrum
is given by power law $P(k)\propto k^{\alpha}$, the structure
functions initially have to be $S_p(r)\propto r^{-p\alpha/2}$. In
the regime of linear evolution, the structure functions will keep to
be $S_p(r)\propto r^{\xi_{l}(p)}$, and the intermittent exponent is
$\xi_{l}(p)=-p\alpha/2$. According to the log-Poisson hierarchy
scenario, the nonlinear evolution leads to the hierarchical transfer
of the power of clustering from large scales to small scales. The
structure function will become $S_p(r)\propto
r^{\xi_{l}(p)+\xi_{nl}(p)}$, where the nonlinear term of the
intermittent exponent is
$\xi_{nl}(p)=-\gamma[p-(1-\beta^{p})/(1-\beta)]$, in which the parameters
$\beta$ and $\gamma$ are dimensionless. $\beta$ measures the intermittency of
the field, and $\gamma$ measures the singularity of the clustering. For Gaussian
field, we have $\beta=1$, and therefore, $\xi_{nl}(p)=0$ for all order $p$.
Since the onset of nonlinear evolution, the parameter $\beta$ will gradually
decrease, and the field becomes intermittent.

With simulation samples, we found that the parameter $\beta$ is decreasing
with the decrease of redshift $z$. It means that the field
is weakly intermittent at earlier time, but strong intermittent at
later time (Liu \& Fang 2008). Although $\xi_{nl}(p)\neq  0$, the
nonlinear evolution keeps the cosmic baryon fluid to be
scale-invariant. We showed that the mass density field of neutral
hydrogen fluid in the scale free range is also well described by the
log-Poisson hierarchy in spite of the neutral hydrogen fraction of
the baryon fluid is not constant in space. This is because the UV
ionization photon is assumed to be uniform, and it does not violate the
scale invariance of this system. However, the number of $\beta$ of
neutral hydrogen is found to be less than that of total baryon
fluid. Therefore, the neutral hydrogen is less intermitted.

The Ly$\alpha$ transmitted flux of quasars Ly$\alpha$ absorption
spectrum is considered to be effective to probe the mass and
velocity fields of neutral hydrogen. However, the redshift
distortion of the velocity field leads to the deviation of the field
of the Ly$\alpha$ transmitted flux from the neutral hydrogen field.
The transmitted flux does not satisfy all predictions of log-Poisson
hierarchy. Fortunately, the effect of radshift distortion is
approximately negligible for some log-Poisson hierarchical predicted
features. Thus, we can test the log-Poisson hierarchy with quasars
Ly$\alpha$ absorption spectrum.

Using high resolution and high S/N data of quasars Ly$\alpha$
absorption spectrum, we show that all the non-Gaussian features
predicted by the log-Poisson hierarchy, including the hierarchical
relation, the intermittent exponent, the ratios of different
moments, and the scale-scale correlation, are consistent with
observed samples. The observed samples of the transmitted flux yield
the same intermittence parameter $\beta$ as that of neutral hydrogen
field produced with hydrodynamic simulation of the concordance
$\Lambda$CDM universe. Our result shows that the intermittent
exponent $\xi(p)$, or parameters $\beta$ and $\gamma$, is effective
to discriminate among models of nonlinear evolution.

The log-normal model can well fit observed data on lower order statistics, but
not good on higher orders. On the other hand, the log-Poisson model gives
good fitting on lower order as well as higher order statistics. Therefore,
a comparison between the log-Poisson model and log-normal model on lower order
statistics will be able to find the relation between parameters of the log-Poisson
and log-normal models. This relations would be useful to explain the parameters
of log-Poisson model with well-known parameters in cosmology, as the parameters
of log-normal model generally are known in cosmology.

Recent studies have shown that the turbulence behavior of
baryon gas can be detected by the Doppler-broadened spectral lines
(Sunyaev et al. 2003; Lazarian \& Pogosyan 2006). Although these
works focus on the turbulence of baryon gas in clusters, the result
is still applicable, at least, for the warm-hot intergalactic medium
(WHIM), which is shown to follow the evolution of Burger's fluid on
large scales (He et al. 2004, 2005). The last but not least, the
polarization of CMB is dependent on the density of electrons, and
therefore, the map of CMB polarization would provide a direct test
on the non-Gaussian features of ionized gas when the data on small
scales become available.

\acknowledgments

We thank Ji-Ren Liu for his contributions in the early stage of
this project. Y. Lu  is supported by China Scholarship Council.
This work was partially supported by US NSF AST 05-07340.

\end{document}